# Toward The Next-Generation Catalyst for Hydrogen Evolution Reaction

Sichen Wei,*[1] Soojung Baek,*[1] Hongyan Yue,[2] Seok Joon Yun,[3] Sehwan Park,[3] Young Hee Lee,[3] Jiong Zhao,[4] Huamin Li,[5] Kristofer Reyes,[1,†] and Fei Yao [1,†]

[1] Department of Materials Design and Innovation, University at Buffalo, Buffalo, New York 14228, USA.

[2] School of Materials Science and Engineering, Harbin University of Science and Technology, Harbin 150040, China.

[3] Center for Integrated Nanostructure Physics, Institute for Basic Science, Department of Energy & Department of Physics, Sungkyunkwan University, Suwon, Gyeonggi-do 16419, South Korea.

[4] Department of Applied Physics, Hong Kong Polytechnic University, Hung Hom, CD 612, Kowloon, Hong Kong

[5] Department of Electrical Engineering, University at Buffalo, Buffalo, New York 14228, USA

----------------------------------

[†] Corresponding authors: kreyes3@buffalo.edu, feiyao@buffalo.edu

**ABSTRACT**

The development of active catalysts for hydrogen evolution reaction (HER) made from low-cost materials constitutes a crucial challenge in the utilization of hydrogen energy. Earth-abundant molybdenum disulfide ($MoS_2$) has been discovered recently with good activity and stability for HER. In this report, we employed the hydrothermal technique for $MoS_2$ synthesis which is a cost-effective and environmentally friendly approach and has the potential for future mass production. To investigate the structure-property relationship, scanning electron microscope (SEM), transmission electron microscope (TEM), X-ray diffraction (XRD), Raman spectroscopy, X-ray photoelectron spectroscopy (XPS), and various electrochemical characterizations have been conducted. A strong correlation between the material structure and the HER performance has been observed. Moreover, machine-learning (ML) techniques were built and subsequently used within a Bayesian Optimization framework to validate the optimal parameter combinations for synthesizing high-quality $MoS_2$ catalyst within the limited parameter space. The model will be able to guide the wet chemical synthesis of $MoS_2$ and produce the most effective HER catalyst eventually.

**KEYWORDS**: Molybdenum Disulfide, Machine-Learning, Hydrogen Evolution Reaction, Bayesian Optimization

**INTRODUCTION**

The ever-increasing demand for energy consumption and harmful $CO_2$ emission necessitate the urgent need for clean energy. As an eco-friendly fuel with the highest gravimetric energy density, the widespread adoption of hydrogen fuel will reduce energy-related emissions and improve energy efficiency[1-4]. Hydrogen can be produced by electrochemical water splitting. Achieving high-efficiency water splitting requires the use of a catalyst to minimize the overpotential to drive the hydrogen evolution reaction (HER) [5-9]. Pt-group metals (PGM) are excellent catalysts for HER, but their practical applications are limited by the high cost and scarcity [10-14]. Therefore, the development of active HER catalysts made from low-cost materials constitutes a crucial challenge in the utilization of hydrogen energy. Recently, earth-abundant two-dimensional (2D) transition metal dichalcogenides (TMDs) have been discovered and demonstrated enormous potentials in various energy-related applications including energy storage, catalysis, electronic devices, and biosensors [15-18]. Among them, molybdenum disulfide ($MoS_2$) has been proved with outstanding catalytic activity and stability for HER [19-23]. $MoS_2$ can be prepared by various techniques including wet chemical synthesis, physical and chemical vapor deposition [24-28]. Nevertheless, the above-mentioned methods normally involve the use of toxic reactants and complicated processes which are not environment friendly and cost-effective.

On the other hand, Machine-Learning (ML) technique has been identified as an effective approach to facilitate material exploration. For example, Kaxiras's group[29] employed a data-driven approach to investigate the magnetic properties of different $A_2B_2X_6$ monolayer structures considering the total energy, magnetic order, magnetic moment, and magnetic excitation energy. Fyta *et. al.* [30] exploited an ML approach to evaluate the lithium adsorption free energy on TMDs which was found to be dominated by the lowest unoccupied state of the substrate. In their report, a set of descriptors were constructed and the adsorption energy of different alkali metals on different TMDs was successfully predicted with the help of ML. Such data-driven methods are amenable when there exists a large body of existing reliable data to which we can fit ML models with a high degree of fidelity. However, in many instances, such data is sparse and expensive to obtain experimentally. Closed-loop design and Bayesian methods, such as Active Learning and Bayesian Optimization (BO)[31] can be utilized to mitigate this problem. BO has been shown effective in the optimization of many materials systems, including optimizing chemical synthesis[32,33], synthesis of quantum dots[34], and phase-change memory nanocomposites[35], to name a few examples. When dealing with a particular device application (such as catalysis) involving novel material development, however, their limitation is exacerbated by the fact that the link to be established, namely that between synthesis parameters to device performance spans the synthesis-material-device scale, meaning statistical correlations inferred between a limited set of synthesis conditions and corresponding device performance metrics are diluted. In addition, multi-objective characterizations of device performance also led to further difficulty in the optimization problem.

Herein, we aim to employ an aqueous-based hydrothermal synthesis technique which is an environmentally friendly and cost-effective approach to achieve high-performance $MoS_2$ HER catalysts. Six sets of initial manually designed synthesis were conducted and one was identified with better HER performance in terms of both overpotential and the Tafel slope. To further improve the HER performance, the synthesis condition needs to be further optimized. Before further exploring the vast parameter space, it is necessary to first validate the current manually picked synthesis parameters represent the local optimality within the explored space in the specific range considering the large parameter-selection intervals. Therefore, we investigated how to use a closed-loop design in an extreme case to perform an extremely limited, criteria-based exploration of synthesis conditions space to validate manually discovered optimized synthesis parameters. Specifically, a Gaussian Process (GP) belief model was built to map the hydrothermal synthesis conditions to a corresponding figure of merit for HER performance. The GP belief model was then used the BO algorithm to suggest the next set of synthesis conditions, which were then used to synthesize a new $MoS_2$ sample. After characterizing the HER activity of the sample, the GP beliefs were updated using Bayesian statistics, and the updated posterior beliefs were used to initiate the next iteration of the closed loop. Through executing this closed loop several times within a limited iteration budget, we can more effectively validate extant optimal conditions than compared to a coarse grained, non-optimization-oriented design under similar experimental budgets. As a result, the BO algorithm validated the optimized $MoS_2$ sample exhibited the best HER performance in terms of overpotential and Tafel slope. The origin of the better HER activity compared to the other samples can be attributed to the fast ion transport associated with the enlarged interlayer space and the increased number of potential active sites for HER originated from the increased surface area.

**EXPERIMENTAL SECTION**

**Materials and Synthesis.** $MoS_2$ samples were synthesized by a hydrothermal technique using ammonium molybdate and thiourea (Fisher Scientific, USA) as precursors. Specifically, ammonium molybdate and thiourea were dispersed in DI water followed by vigorously stirring until the solids were completely dissolved and a transparent solution was obtained. The solution was then transferred into a 100 ml Teflon line autoclave and heated in a convection oven. The reaction conditions were manually selected for the initial design of synthesis. Specifically, the temperature was set between 180 and 205 °C. The reaction time was varied from 16 to 26 hours. The Mo and S precursor concentrations were selected between 0.02 ~ 0.03 mol/L and 0.65 to 0.9 mol/L with a step variation of 0.002 and 0.05 mol/L, respectively. The solution was then naturally cooled down to room temperature after reaction and the precipitate was collected by centrifugation and washed using DI water and ethanol for three times each. The final product was then dried in a vacuum oven overnight at 60 °C. The details for manually designed synthesis conditions are listed in **Table S1**.

**Characterization.** The morphology and composition of all products were investigated by field-emission scanning electron microscope (FE-SEM) and energy dispersive X-ray (EDS) (Carl Zeiss AURIGA CrossBeam with Oxford EDS system). The transmission electron microscopy (TEM) was conducted using the JEM ARM 200F system. X-ray diffraction and Raman spectroscopy were performed using a Rigaku Ultima IV with Cu Kα radiation (wavelength at 1.541 nm) and Renishaw InVia with an excitation laser wavelength of 514 nm, respectively. X-ray photoelectron spectroscopy (XPS) is performed using a monochromatic Al Kα source (hv = 1486.6 eV, ESCALAB 250, Thermo Scientific). Brunauer-Emmett-Teller (BET) specific surface area was measured on the Micromeritics Tri-Star II system by nitrogen ($N_2$) adsorption-desorption isotherm at 77 K.

**Electrochemical Measurement.** The ink for the HER test was prepared by dissolving 10 mg of as-prepared $MoS_2$ in a mixture of 500 µL of ethanol, 500 µL of DI water, and 15 µL Nafion D-521 solution. The electrochemical characterization was performed using CHI760E electrochemical workstation (CH Instrument) in a standard three-electrode configuration which consists of a silver/silver chloride (Ag/AgCl in 1 M KCl), a platinum (Pt) wire, and an ink-coated glassy carbon rotating ring disc electrode as reference, counter and working electrodes, respectively. The reference electrode was converted to the potential vs. $E_{RHE}$ based on the equation: $E_{RHE} = E_{Ag/AgCl} + 0.059\ pH + E°_{Ag/AgCl}$, where $E°_{Ag/AgCl} = 0.222$ V. The loading amount of $MoS_2$ catalyst is 0.285 mg cm$^{-1}$ and the samples were cycled 20 times before any data recording. Nitrogen gas saturated 0.5 M H$_2$SO$_4$ was employed as electrolyte. The linear sweep voltammetry (LSV) was carried out at a scan rate of 5 mV/s and the built-in IR compensation was executed prior to LSV tests. The electrochemical impedance spectroscopy (EIS) was tested from 0.1 Hz to 1M Hz at an overpotential of 250 mV. For each sample, the cyclic voltammetry (CV) was carried out in a series of scan rates (20, 40, 50, 60, 80, and 100 mV/s) in the potential range of 0.05 ~ 0.15 V versus RHE. The double-layer capacitance (C$_{dl}$) was assessed from the slope of the linear regression between the current density differences ($\Delta$J/2=(J$_{anode}$-J$_{cathode}$)/2 at an overpotential of 0.1 V vs. RHE) versus the scan rates. The accessible surface area of as-synthesized samples could be approximated from the electrochemical active surface area (ECSA). The ECSA was determined by ECSA=C$_{dl}$/C$_s$, where C$_s$ stands for the specific capacitance of standard electrode materials on a unit surface area. Here, based on the literature reported C$_s$ values for flat surfaces, 0.04 mF/cm$^2$ was used for ECSA calculations[36]. The turnover frequency (TOF) was calculated from CV measurement. The CV curve was measured in phosphate buffer solution (pH = 7) at a scan rate of 50 mV/s. The TOF was then calculated following the equation: *TOF=I/2Fn*, where *I* is the current from the polarization curve, *F* is the Faradic constant and *n* is the number of active sites. Among them, the *n* can be estimated using the relationship: *n=Q/2F*, where *Q* is the voltammetric charges and *F* is the Faradic constant, respectively.

**Machine Learning Theory.** Closed-loop Bayesian exploration of synthesis conditions was performed to validate the optimality of the manually selected model within a limited experimental budget. Six batches of MoS$_2$ synthesis conditions (Mo and S precursor amount, temperature, and reaction time) and the corresponding HER activity (evaluated by overpotential at the current level of 10 mA/cm$^2$, *i.e.,* η10, and the Tafel slope) were used (see **Table S1**) to train Gaussian Process (GP) belief model to express a scalar combination of the electrochemical responses as a function of input synthesis conditions. Implicit in this modeling is that the material precursors, apart from the concentration, remain constant throughout the campaign. This technique can be adapted to learning optimal synthesis conditions and choice of material, given an appropriate parameterization of material properties as additional features or inputs to the ML model. The GP was fit to these first six "seed" data points, resulting in a Bayesian prior GP, $B^6$. We then selected experiments based on this prior, according to one of several decision-making policies such as the Expected Improvement (EI), Upper Confidence Bound (UCB), Maximum Variance (MV), Exploration, and Exploitation policies. A full derivation of the policies is provided in the Supplemental Information. Broadly speaking, EI, UCB, and Exploitation policies are those geared toward optimization, while MV and Exploration are policies meant for generally learning the response function. In general, the policies would score potential experiments based on GP beliefs and some measure of information gain. The experiments with maximal scores were selected. We refer the reader to the Supplementary Information for more details.

Prior to running the physical, closed-loop experimental campaign, we performed simulations of this campaign to assess model and policy performance for the given problem. From the prior $B^6$, we sampled a surrogate for the ground truth response function $f_i^* \sim B^6$. We then selected an experiment according to some policy $\mathrm{x}^7 = \mathrm{argmax}\, a(\mathrm{x}; B^6)$, and then simulated a noisy observation of the ground truth:

$$y_7 = f_i^*(\mathrm{x}^7) + W,$$

where W is sampled from $N(0;\ \sigma_W^2)$. This is used to form the simulated posterior belief $B^7$, which was subsequently treated as the prior belief for the next iteration of the closed loop. By iterating through this loop several times, we simulated an experimental campaign. We then simulated 100 such campaigns, each with different surrogates of the ground truth, $f_1^*, \ldots, f_{100}^*$. We then aggregated simulation results to calculate a performance metric called relative Opportunity Cost (OC), which is a relative measure of how optimal the predicted optimum synthesis conditions are compared to the true optimal conditions, with smaller OC values implying more optimal predictions.

**RESULTS AND DISCUSSION**

**Initial manual design of synthesis and performance evaluation for aqueous-based MoS₂ catalyst.** We employed a hydrothermal technique for MoS₂ synthesis where ammonium molybdate and thiourea were used as Mo and S precursors, respectively. Such a method involves no use of toxic reactants and therefore has a potential for large-scale adoption due to its environmentally friendly and cost-effective nature. The initial design of synthesis conditions was manually picked based on our experimental experiences in hydrothermal synthesis and each parameter was confined in a specific range, as shown in **Table S1**. Six batches of manually designed samples were labeled as MoS₂-1 to MoS₂-6 and the corresponding HER activities were evaluated by LSV measurements, as shown in **Figure 1(a)**. The values of η10 were extracted from the polarization curves which are 288, 243, 240, 254, 254, and 249 mV for sample MoS₂-1 to MoS₂-6, respectively (see **Table S1**). It is clear to see that MoS₂-3 exhibited the smallest η10 value, indicating a better HER activity compared to the rest of the manually designed samples.

As a multistep reaction, HER starts with Volmer reaction ($H_3O^+ + e^- \rightarrow H^* + H_2O$) and the intermediate (adsorbed H*) is desorbed from the catalyst surface by either Tafel reaction ($H^* + H^* \rightarrow H_2$) or Heyrovsky reaction ($H_3O^+ + H^* + e^- \rightarrow H_2 + H_2O$) in acidic media. The values of the Tafel slope can be used to identify the rate-determine step [37,38]. The Tafel slopes are plotted in **Figure 1(b)** and the related values are summarized in **Table S1**. Except for MoS$_2$-1 which showed a poor HER performance, all of the other samples showed the Tafel slope value in the range of 60 ~ 80 mV/dec, suggesting that Heyrovsky reaction was the rate-determine step. The slow Heyrovsky reaction can be ascribed to the large MoS$_2$ resistance resulting in a limited number of electrons that can be transferred to the catalyst/electrolyte interface. The EIS measurement was carried out to investigate the charge transfer resistance. The resulting Nyquist plots of each sample are shown in **Figure 1(c)**, where the charge transfer resistance ($R_{ct}$) governed HER kinetics can be evaluated by the radius of the semicircle. Smaller $R_{ct}$ values were observed for MoS$_2$-3 (31.0 Ω) and MoS$_2$-6 (48.8 Ω). Considering the values of both overpotential and Tafel slope, we observed generally better HER performance for MoS$_2$-3 among the manually designed 6 synthesis conditions.

**Prior policy assessment through simulations.** Since the initial 6 batches of synthesis were fully determined by the experimentalists' experience and intuition, potentially superior synthesis conditions may exist but were not chosen by the experimentalists in the initial manual design process. To evaluate the optimality of MoS$_2$-3, we performed a limited BO to discover potentially superior synthesis conditions. In BO, decision-making policies allocate a small number of experiments between exploring the parameter space and focusing on regions believed to yield promising results, based on a limited understanding of the landscape it is exploring. This is called the exploration versus exploitation trade-off, and policies that perform this balance well typically are able to identify optima in fewer experiments than trial-and-error or grid-based search approaches [39,40].

In the past, BO has been used to efficiently explore synthesis parameter space in order to optimize chemical and material properties [34,41]. In these past applications, the BO was executed on a relatively large number (50 - 100 s) of experiments executed by autonomous platforms. In this current setting, however, due to the non-autonomous, manual work and time needed to execute the synthesis and characterization of electrochemical performance, validation of $MoS_2$-3 is limited to a small number (less than 10) of experiments. In addition, the characterization spans from synthesis and processing parameters of material to device performance, requiring the optimization of several electrochemical quantities.

In this limited application, we utilized BO techniques to simply evaluate the optimality of $MoS_2$-3 with the goal of exploring parameter space in an objective-driven manner as much as possible. That is, given the prior results provided by $MoS_2$-3, we wish to utilize BO methods as a systematic way of validating its quality given the limited experimental budget, in contrast to further *ad hoc* search. We performed a Bayesian sequential design of experiments using GP belief models and the decision-making policies outlined in the above section. GP beliefs were fit to the initial seed data set of 6 synthesis conditions, and a decision-making policy would select an experiment to run based on these beliefs. Once the experiment was run, electrochemical performance was evaluated, the results of which were used to update the GP beliefs. This process would be repeated until the small experimental budget was expended. To select a specific decision-making policy to use during this validation campaign, we ran several simulations of the campaign. The use of such statistical simulation to perform meta-decision-making regarding policy and modeling choices has been shown effective in the past [42–44]. We calculated OC curves for all four policies, which are shown in **Figure S1**. From this simulation study, we observed that the UCB policy had the best simulated performance.

**Parameter Space Exploration.** To quantify the effectiveness of the ML-guided closed-loop exploration campaign in exploring synthesis parameter space, we calculated a data-spread metric. Specifically, for a set of data points $X = \{x_i\}$, we calculate the mean distance from the centroid

$$s(X) = \frac{1}{|X|}\sum_{i=1}^{|X|}\|x_i - \mu(X)\|_2,$$

where $\mu(X)$ is the centroid of the data points $X$ and it is a measure of the spread of the data.

$$\mu(X) = \frac{1}{|X|}\sum_{i=1}^{|X|} x_i.$$

**Figure S2** plots the distribution of this spread over simulated experimental campaigns using the UCB, EI, and MV policies. In addition, this statistic was calculated for the actual campaign data, $s_{\text{data}}) = 0:34$. We plot this value as a vertical red dashed line. Examining the plots, we show that the spread for the actual data is larger than a majority of the simulated spreads using both the EI and UCB policies, indicating that the experimental campaign was able to explore well within the limited set of experiments available to it. As expected, the MV policy does explore more, as indicated by the larger spread values. This, along with the simulation results in **Figure S1** indicate that balanced decision-making policies such as EI and UCB consistently outperform more exploration-oriented policies such as MV in finding optima within a small number of experiments.

The sequential study based on BO predictions was performed and after 10 steps of the study, a final predicted condition was given. The samples were synthesized correspondingly and the synthesis conditions for each step can be found in **Table S1**. Surprisingly, we found that the synthesis condition predicted by the BO algorithm for the optimized $MoS_2$ catalyst (labeled as $MoS_2$-Opt) is very similar to that of $MoS_2$-3, which has been identified as a better HER catalyst compared to the other samples synthesized by initial manually selected conditions. As a result, the BO algorithm has determined that a synthesis condition close to MoS$_2$-3 can yield the promising HER activity after the step-wise study.

The campaign shows the viability of such techniques in systematically verifying the performance of MoS$_2$-3 in a limited experimental setting. BO is one of the limited techniques that can operate in this few-shot environment[45]. This process can be accelerated to some degree through the inclusion of informative priors for the GP model, which can come from related systems[46] or elicited directly from domain experts[47]. By augmenting the above procedure with such methods, we can overcome the experimental limitations for a more effective search.

**Experimental Validation.** To validate the BO algorithm predictions, HER activities of different MoS$_2$ samples synthesized at ML-predicted conditions (labeled as MoS$_2$-Px, where Px indicates the predicated batch number) were investigated using electrochemical techniques and the related results were summarized in **Table S1.** HER performance for selected representative samples (i.e. MoS$_2$-P3, MoS$_2$-P9, and MoS$_2$-Opt) is shown in **Figure 2**. In **Figure 2(a)**, the MoS$_2$-Opt exhibits a small η10 of 240 mV and a Tafel slope of 64 mV/dec, which is very close to that of MoS$_2$-3 (243 mV, 70 mV/dec), demonstrating good reproducibility of these two synthesis conditions. The overpotentials and Tafel slopes for MoS$_2$-P3 and MoS$_2$-P9 were 219 mV, 88 mV/dec, and 253 mV, 72 mV/dec, respectively. Considering the HER performance is evaluated in terms of both overpotentials and Tafel slopes, the overall HER activity of MoS$_2$-Opt outperformed the rest of the ML-predicted samples. Although MoS$_2$-P3 has a smaller η10, a larger Tafel slope compared to MoS$_2$-Opt makes it unfavorable as an efficient HER electrocatalyst. Furthermore, MoS$_2$-Opt displayed a smaller R$_{ct}$ of 49 Ω compared to that of MoS$_2$-P3 (60 Ω) and MoS$_2$-P9 (50 Ω) (see **Figure 2(c)**), suggesting its fast charge transport at the catalyst/electrolyte interface and therefore a good HER activity. In addition, the polarization curves of the MoS$_2$-Opt before and after 1000

CV cycles at a scan rate of 50 mV/s showed negligible degradation as shown in **Figure 2(d)**, demonstrating outstanding long-term stability of the BO-predicted optimized $MoS_2$ sample.

To gain insight into the origin of the better HER performance in the BO-predicted $MoS_2$-Opt, a series of characterizations were carried out. The morphology of the as-synthesized $MoS_2$-Opt was investigated using SEM and TEM. As shown in **Figure 3(a)**, the $MoS_2$-Opt exhibited a nanoflower-like structure with the tendency to form aggregated bundles. The TEM image shown in **Figure 3(b)** confirmed the flower-shaped $MoS_2$ morphology assembled by wrinkled $MoS_2$ nanoflakes. The HRTEM analysis shown in **Figure 3(c)** revealed that the typical $MoS_2$-Opt showed polycrystalline structure as evidenced by the clear ring patterns in the SAED pattern. Furthermore, 2H phase $MoS_2$ with trigonal prismatic coordination was observed, as shown in **Figure 3(d)**. The corresponding EDS elemental mapping of Mo, S, and O in **Figure 3(e-h)** revealed a uniform distribution of Mo and S, indicating the successful formation of $MoS_2$. The emerging of oxygen signals in the EDS mapping can be ascribed to the sample oxidation after preparation.

It is well known that the HER performance of 2H phase $MoS_2$ is inferior to that of 1T $MoS_2$ due to its semiconducting nature. The relatively large electrical resistance in 2H $MoS_2$ will hinder the electron transport to catalyst/electrolyte interface and thus large $R_{ct}$ values and slow Heyrovsky reaction can be expected. This eventually leads to relatively poor HER catalytic activity in all of our aqueous-based samples. Nevertheless, the introduction of metallic 1T $MoS_2$ normally requires toxic organic reactants and complicated process control [48–50], which re-emphasizes the importance of the current work with the focus of achieving a high-performance, PGM-free HER catalyst via a green synthesis process.

The chemical composition of the optimized sample was investigated using XPS, as shown in **Figure 4(a)**. The complete survey spectrum of $MoS_2$-Opt showed the typical $MoS_2$ bonding information with characteristic signals for Mo 3d and S 2p, which is consistent with the previous reports [51,52]. The O 1s peak was also observed due to the unavoidable oxidation of the sample. Two major peaks located at ~ 228.5 and 231.7 eV can be found in Mo 3d spectrum in **Figure 4(b)** which can be assigned to the $Mo^{4+}$ $3d_{5/2}$ and $Mo^{4+}$ $3d_{3/2}$, respectively, with an S 2s peak at ~ 225.6 eV. As for the S 2p spectrum in **Figure 4(c)**, doublet peaks are found at ~ 161.3 and 162.4 eV corresponding to the S $2p_{3/2}$ and S $2p_{1/2}$ peaks, respectively. The Mo 3d and S 2p peaks for all the samples correspond well with the previous reports [51,52], proving the successful formation of $MoS_2$. In addition, the deconvoluted spectrum for S 2p showed one set of doublet peaks, indicating the as-prepared $MoS_2$ contained only 2H phase, resonating well with our TEM observations [53,54]. The O 1s spectrum shown in **Figure 4(d)** showed two peaks at 531 and 532 eV resulted from the bonding of Mo-O and adsorbed water, indicating the slight oxidation of the $MoS_2$ sample. XPS spectrum for $MoS_2$-P3 and $MoS_2$-P9 were shown in **Figure S3** and **Figure S4** and no obvious differences were found compared to $MoS_2$-Opt.

The crystal nature and interlayer spacing of different $MoS_2$ samples synthesized at ML-predicted conditions were investigated by XRD, as shown in **Figure 4(e)**. The peaks at 14.6°, 32.3°, and 57.3° can be ascribed to (002), (100), and (110) crystal planes of the 2H phase, respectively, which are consistent with previous reports for $MoS_2$ [55]. For $MoS_2$-P9 and $MoS_2$-Opt, a downshift of (002) peak from ~ 14 ° to ~ 9 ° along with the emerging of a new peak at 18° which corresponds to the (004) crystal plane was observed indicating an enlarged interlayer spacing [56]. The extracted interlayer distance for $MoS_2$-Opt, P9, and P3 are 0.94, 0.92, and 0.64 nm, respectively (see **Table S2**). The highest interlayer spacing value in $MoS_2$-Opt will not only provide a large surface area but also benefit ion diffusion for the HER process, which is the main reason for the observed better HER activity as shown in **Figure 2**. Besides that, the average crystallite size was also extracted from the XRD (002) peaks based on the Scherrer Equation. As a result, the average crystallite sizes of $MoS_2$-P3, $MoS_2$-P9, and $MoS_2$-Opt were calculated to be 1.97 nm, 6.4 nm, and 6.94 nm, respectively, which is consistent with the previously reported value[57]. Furthermore, typical $E_{2g}$ and $A_{1g}$ vibration peaks were observed in the Raman spectrum for all samples, as can be seen in **Figure 4(f)**. The result indicated that the as-synthesized aqueous-based $MoS_2$ samples are solely 2H phase which is in line with the TEM and XPS observations in **Figure 3(d) and Figure 4(c)**. Brunauer-Emmett-Teller (BET) analysis in **Figure S5** showed that $MoS_2$-Opt exhibited a larger specific surface area of 23.9 $m^2$ $g^{-1}$ compared to the other ML-predicted samples (see **Table S3**) which can be attributed to the enlarged interlayer spacing as observed in **Figure 4(e)**. Furthermore, the pore size distribution was extracted by the Barrett-Joyner-Halenda method (BJH), as can be seen in **Figure S5(b)**. All samples exhibit a sharp peak at ~ 5 nm and a broad peak at a range of 20 ~ 40 nm, indicating a mesopore-enriched structure of as-prepared $MoS_2$. The large surface area associated with abundant mesopores and large interlayer distance will facilitate the ion diffusion toward the catalyst surface, leading to an enhanced HER activity. The accessible surface area of as-synthesized samples was evaluated by ECSA, as shown in **Figure S6**. The $C_{dl}$ and the corresponding ECSA were summarized and listed in **Table S4**. The extracted ECSA of $MoS_2$-Opt is 929.1 $cm^2$, which is the largest among all samples and resonates well with our observation in **Figure S5**. The Turnover frequency of $MoS_2$-Opt was also extracted to be 0.042 $H_2$/s at an overpotential of 100 mV, as can be seen in **Figure S7**, which is corresponding to the previously reported value for $MoS_2$[20,58]. Overall, the origin of the better HER performance observed in the $MoS_2$-Opt sample can be attributed to the increased interlayer space which facilitated the charge transfer at the electrode/electrolyte interface, and the enlarged surface area which increased the number of potentially HER active sites.

**CONCLUSION**

In summary, we employed an aqueous-based hydrothermal technique to synthesis $MoS_2$ as an alternative catalyst of PGM for hydrogen production. This method does not involve any toxic reactants or gases which is environmentally friendly and cost-effective. We used BO to validate and optimize the manually obtained high-performing $MoS_2$ sample. It is important to note that an exhaustive exploration of the parameter space is not the goal of BO. Instead, BO attempts to identify optimal parameters through a combination of exploration and exploitation. Within a small set of 9 validation experiments selected through the UCB policy, we identified an optimized set of synthesis conditions ($MoS_2$-Opt) that, while resulting in better overall performance, are similar to those conditions in $MoS_2$-3. Note that this result does not imply that $MoS_2$-3 is globally optimal, which would require a larger set of experiments to be performed. Instead, the use of BO was meant to validate its optimality as much as possible within an extremely limited experimental budget, pointing to the potential of bootstrapping such ML methods to work, validate and improve manually identified synthesis conditions. Moreover, the catalysis performance of as-synthesized $MoS_2$ is still inferior to the solvothermal synthesized $MoS_2$ due to the existence of pure 2H phase. To further improve the HER performance, structure engineering strategies including defect introduction, phase engineering, and composite formation will be utilized in the future.

**ASSOCIATED CONTENT**

**Supporting Information**

**AUTHOR INFORMATION**

**Corresponding authors**

kreyes3@buffalo.edu, feiyao@buffalo.edu

**Author contributions**

K. R. and F. Y. conceived and supervised the project. S. W. and H. Y. Yue synthesized and performed electrochemical characterizations. S. B. completed the machine-learning simulations and calculations. S. J. Yun, S. W. P., J. Zhao, Y. H. Lee, and H. M. Li performed materials characterization and analysis. All authors discussed the results and contributed to the final manuscript.

**Acknowledgments**

This work was partially supported by New York State Energy Research and Development Authority (NYSERDA) under Award 138126, and the New York State Center of Excellence in Materials Informatics (CMI) under Award C160186. The authors acknowledge support from the Vice President for Research and Economic Development (VPRED) at the University at Buffalo.

**Ethics declarations**

The authors declare no competing interests.

**FIGURES**

**Table of Content**

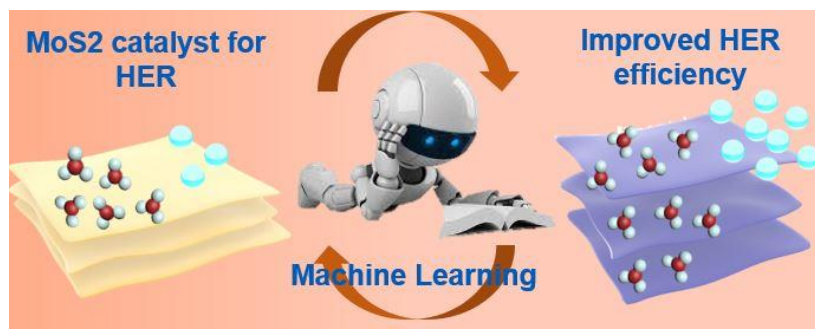

The schematic illustration of machine-learning technique assisted high-performance electrocatalyst development.

**Figures in Manuscript**

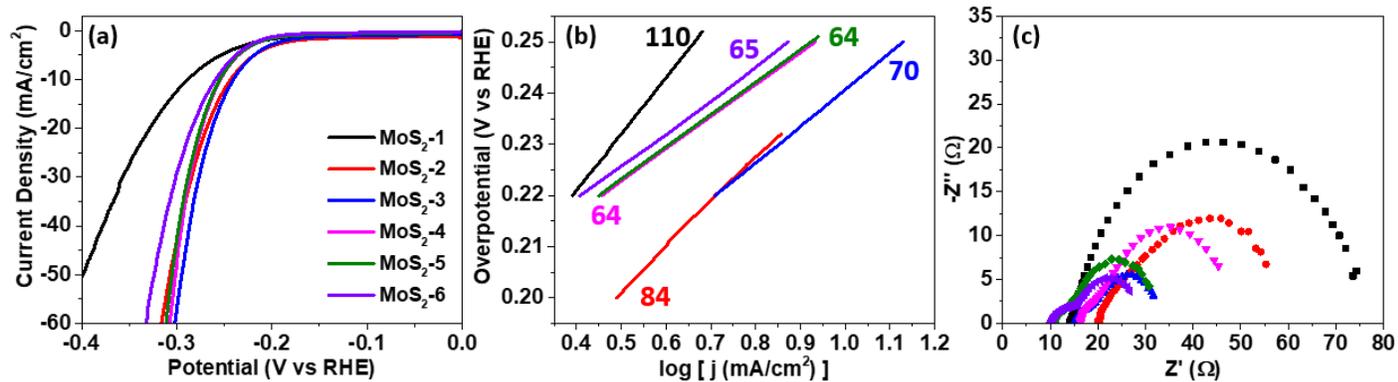

**Figure. 1**. (a) The polarization curves, (b) the Tafel slopes, and (c) the Nyquist plot of the first 6 batches of manually designed MoS$_2$ samples.

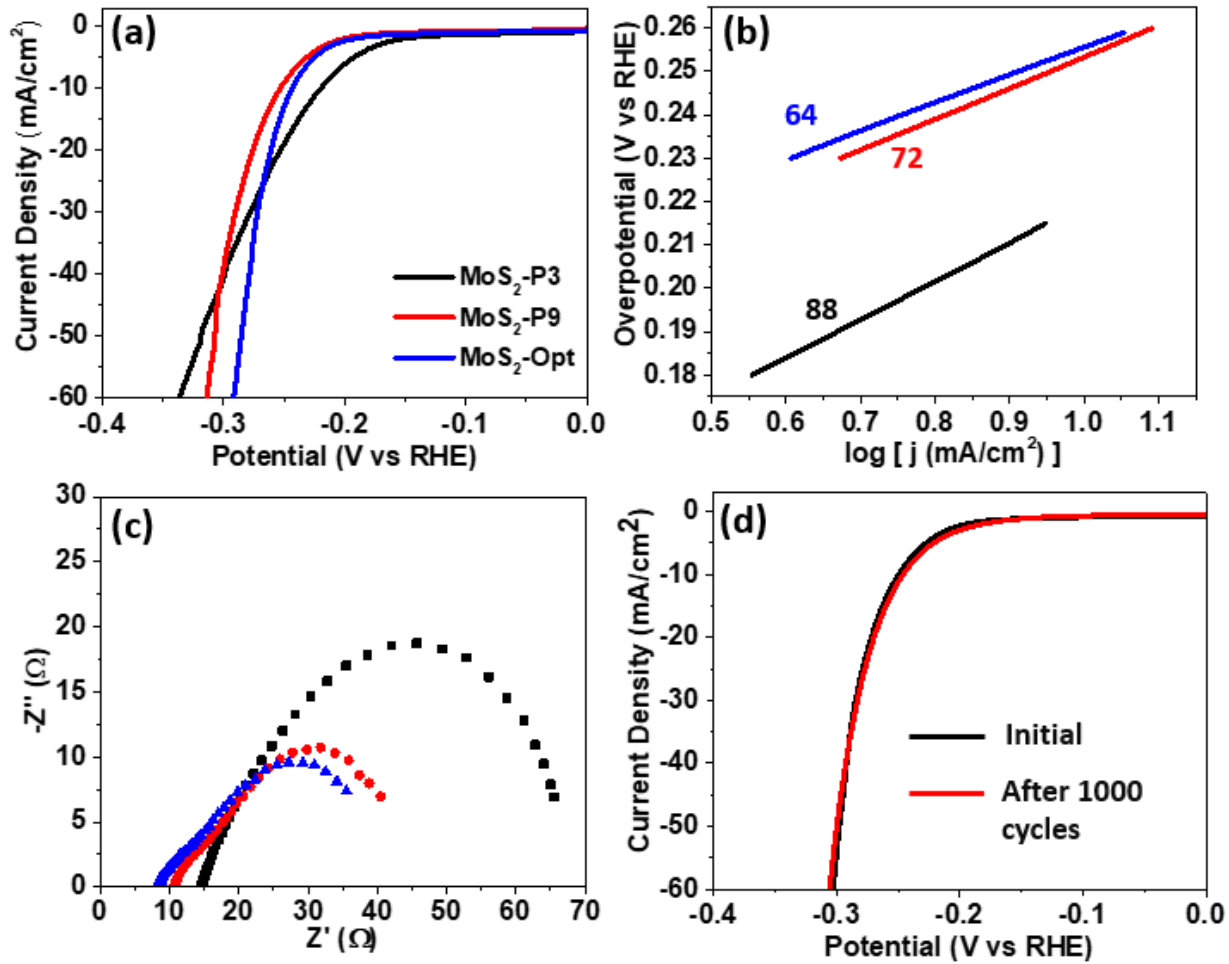

**Figure. 2.** (a) Polarization curves, (b) Tafel plots, and (c) Nyquist plots of the selected ML-predicted samples. (d) The polarization curves of MoS$_2$-Opt before and after 1000 cycles of CV scan.

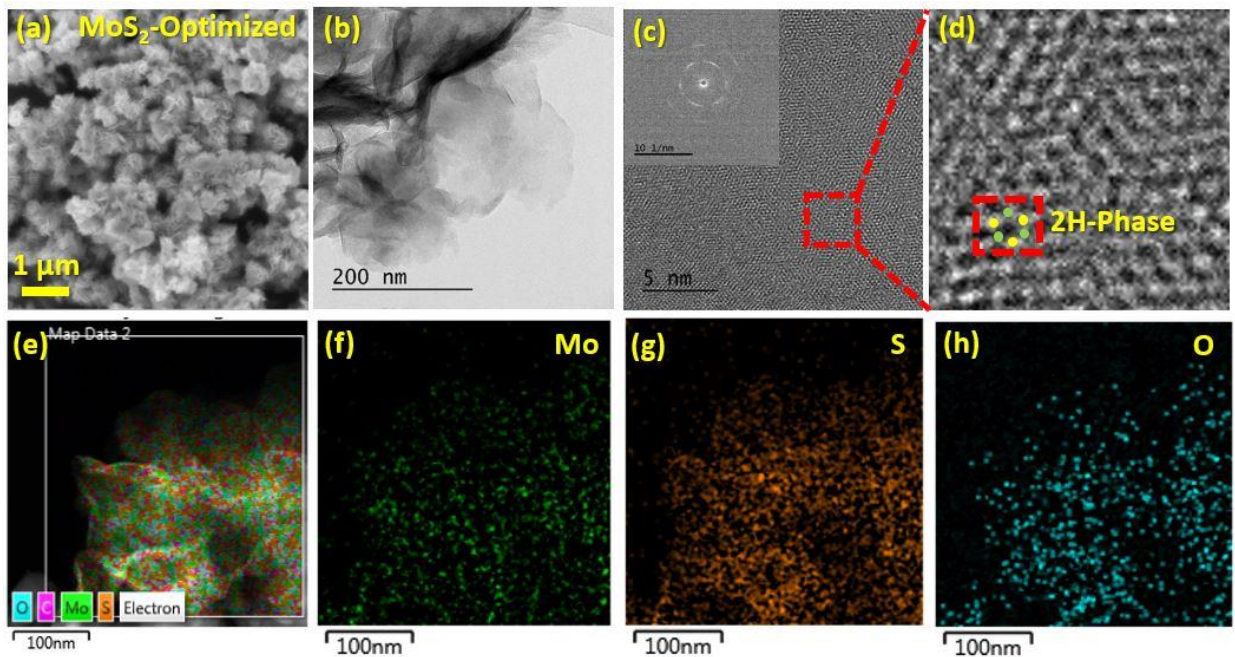

**Figure. 3.** (a) SEM and TEM images for the MoS$_2$-Opt sample. (c) HRTEM image and corresponding SEAD pattern. (d) HRTEM image of the enlarged area in (c). The 2H phase atomic arrangement was marked in color. (e)-(h) EDS mapping of Mo, S, and O.

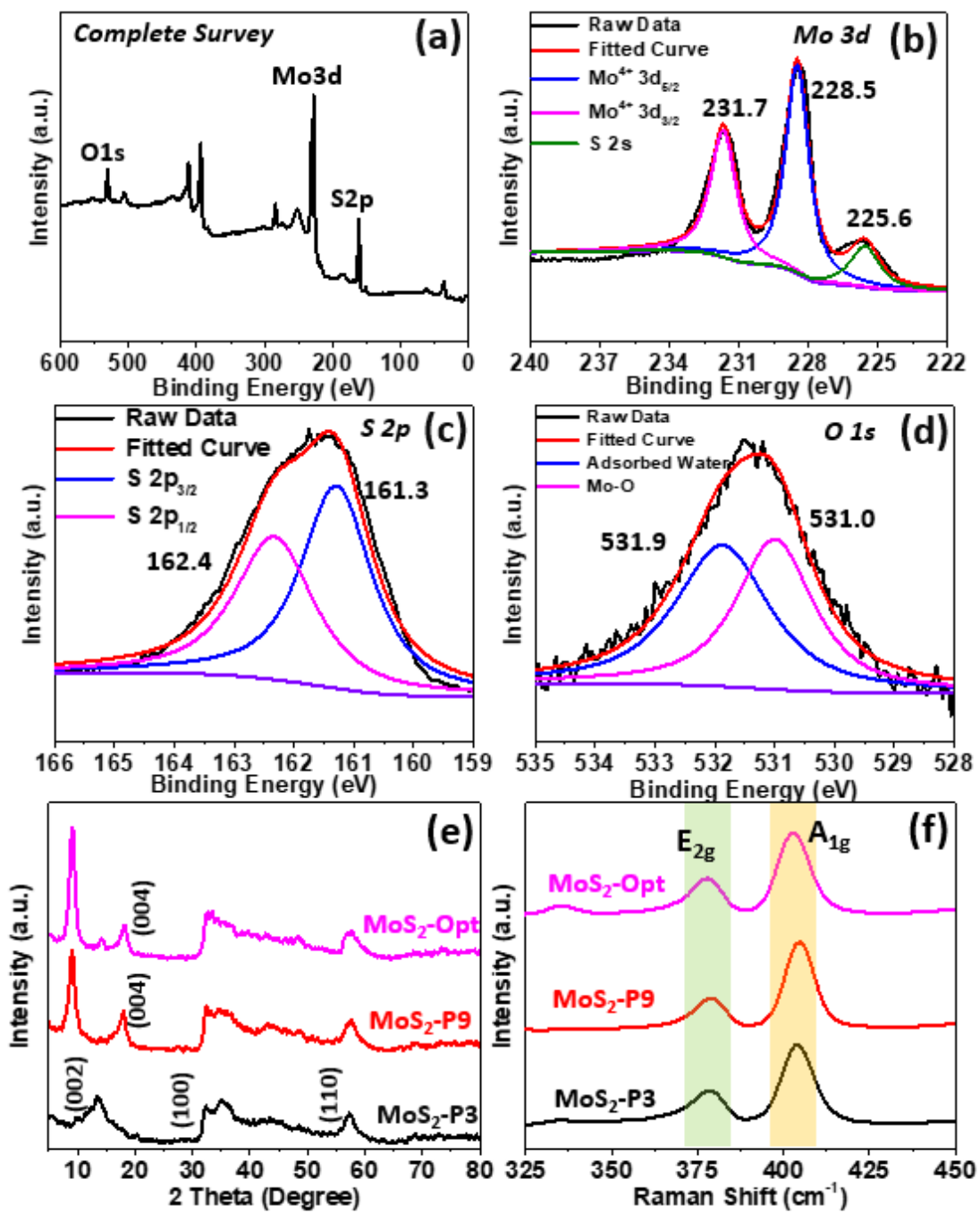

**Figure. 4.** (a) The complete XPS survey spectrum for MoS$_2$-Opt, (b) the Mo 3d spectrum, (c) the S 2p spectrum, and (d) the O 1s spectrum. (e) The powder XRD pattern, (f) Raman spectra.

# Supplementary Information

**Toward The Next-Generation Catalyst for Hydrogen Evolution Reaction**

Sichen Wei [1, †], Soojung Baek [1, †], Hongyan Yue [2], Seok Joon Yun [3], Sehwan Park [3], Younghee Lee [3], Jiong Zhao [4], Huamin Li [5], Kristofer Reyes [1, *] and Fei Yao [1, *]

[1] *Department of Materials Design and Innovation, University at Buffalo, Buffalo, New York 14228, USA*

[2] *School of Materials Science and Engineering, Harbin University of Science and Technology, Harbin 150040, China*

[3] *Department of Energy & Department of Physics, Sungkyunkwan University, Suwon, Gyeonggi-do, South Korea*

[4] *Department of Applied Physics, Hong Kong Polytechnic University, Hung Hom, Kowloon, Hong Kong*

[5] *Department of Electrical Engineering, University at Buffalo, Buffalo, New York 14228, USA*

*Corresponding author email: feiyao@buffalo.edu; kreyes3@buffalo.edu

**Machine Learning Models and Decision-Making Policies**

The machine learning that was used in this study comprises of two main parts. The first is the development of a Gaussian Process (GP) model to represent the estimate of the objective function $f^*(x)$ as a function of synthesis conditions **x**. Here,

$$\mathbf{x} = (x_{Mo}, x_S, x_{temp}, x_{time})$$

is a four-dimensional vector whose entries respectively correspond to the concentration of the Mo precursor, the concentration of the S precursor, temperature and time. To account for differing units, raw data describing synthesis was non-dimensionalized using a linear transformation. For example, a raw temperature quantity $T$ was non-dimensionalized according to:

$$x_{temp} = \frac{T - T_{min}}{T_{max} - T_{min}}.$$

where $T_{max}$ = 180°C and $T_{min}$ = 220°C are pre-specified temperature bounds. Similar transformations and bounds were used for the three other synthesis conditions. After $n$ experiments, we view the objective function $f^*$ as a random function with distribution described by a GP:

$$f^* \sim GP(\mu^n(\mathbf{x}), \Sigma^n(\mathbf{x}, \mathbf{x}')),$$

which is described by a mean estimate $\mu^n$ of $f^*$ and a covariance function $\Sigma^n$. Before any data is obtained, we specify a GP prior distribution:

$$\mu^0(\mathbf{x}) = \mu_0 \tag{1}$$

$$\Sigma^n(\mathbf{x}, \mathbf{x}') = \sigma_0 exp\left[-\frac{\|\mathbf{x}-\mathbf{x}'\|_2^2}{2\ell^2}\right]$$

$$\tag{2}$$

where $\mu_0 = 0.5$, $\sigma_0 = 0.5$ and $\ell = 0.2$ are parameters chosen prior to any experiments being run. Upon observing data $D = \{(x_i, y_i)\}_{i=1}^n$, we assume $y_i$ are noisy observations

$$y_i = f^*(x_i) + W$$

where $W \sim N(0, \sigma_W^2)$ is additive Gaussian noise, with variance $\sigma_W^2$. Throughout, we assume $\sigma_W^2 = 0.1^2$. Given this data, the GP posterior distribution $GP(\mu^n, \Sigma^n)$ is derived from both the data and prior:

$$\mu^n(x) = K(x, X^n)[K(X^n, X^n) + \sigma_W^2 I]^{-1} y^n \qquad (3)$$

$$\Sigma^n(x, x') = K(x, x') - K(x, X^n)[K(X^n, X^n) + \sigma_W^2 I]^{-1} K(X^n, x')$$

$$(4)$$

where $X^n$ is the data matrix whose *i*-th row is $x_i$, $y^n$ is the column vector whose *i*-th entry is $y_i$, and for any two data matrices *A* and *B* with *m* and *n* rows, respectively, *K(A, B)* is an $m \times n$ matrix of covariance values whose *(i, j)*-the entry is given by

$$K(A, B)_{i,j} = \sum{}^0 (a_i, b_j)$$

where $a_i$ and $b_j$ are is the *i*-th row of *A* and *j*-th row of *B*, respectively.

The use of GP to model the response surface $f^*$ continuously was due to the dimensionality of the problem. Such a continuous method could result in arbitrarily small differences between experimental inputs, which may be undesirable. This can be addressed with various heuristics, to avoid areas previously explored. Such heuristics employ augmenting length-scale and noise hyperparameters of the GP model, or explicitly multiplying the acquisition function by δ- or Gaussian functions to decrease the perceived utility of an experiment nearby previous experiments. Such methods augment the balance between exploration and exploitation inherent stricken in decision-making policies such as UCB and EI, and often require the calibration of further hyperparameters.

Under time-$n$ beliefs $B^n = GP(\mu^n; \Sigma^n)$ on $f^*$ described by equations (3) and (4), an informationally optimal experiment $x^{n+1}$ experiment can be selected according to a decision-making policy. Several policies were considered. The Expected Improvement (EI) policy selects the next experiment $x^{n+1}$ as the one that maximizes the EI acquisition function

$$a_{EI}(x; B^n) = \mathbb{E}[[f^*(x) - y_{best}^n]^+],$$

Where $y_{best}^n = \max x_i \leq ny_i$ and

$$[c]^+ = f(x) = \begin{cases} c, & c \geq 0 \\ 0, & otherwise \end{cases}.$$

Here, the expectation $\mathbb{E}$ taken over the current beliefs $B^n$ for the unknown response function $f^*$. Integrating this expression yields

$$a_{EI}(x; B^n) = [\Delta^n(x)]^+ + \sigma^n(x)\varphi\left(\frac{\Delta^n(x)}{\sigma^n(x)}\right) - |\Delta^n(x)|\Phi\left(\frac{\Delta^n(x)}{\sigma^n(x)}\right),$$

where $\Delta^n(x) = \mu^n(x) - y_{best}^n$ and $\sigma^n(x) = \Sigma^n(x, x)$. Here, $\varphi(\cdot)$ and $\Phi(\cdot)$ are the standard normal probability density and cumulative distribution functions, respectively. The next experiment is selected through an optimization

$$x^{n+1} = \underset{\chi}{\operatorname{argmax}} \, a_{EI}(x; B^n),$$

where the optimization was done over the 4-dimensional hypercube $\chi = [0, 1]^4$.

Another policy is the Upper Confidence Bound (UCB) policy, which selects the experiment based on the acquisition function:

$$a_{UCB}(x; B^n) = \mu^n(x) + c\sqrt{\Sigma^n(x, x)},$$

where $c$ is a user-specified parameter, here set to $c = \sqrt{2}$. Two baseline policies were also considered: the Maximum Variance (MV) policy, with acquisition function:

$$a_{MV}(x; B^n) = \Sigma^n(x, x),$$

and the Pure Exploitation policy (XPLT), with acquisition function:

$$a_{XPLT}(x; B^n) = \mu^n(x).$$

**Simulations and Opportunity Costs**

During the $n$-th iteration of the $i$-th simulation campaign, we can calculate the predicted optimum design $x_i^{n,*}$ based on the beliefs $B^n$. We evaluate the ground truth at this predicted optimum, $f_i^*(x^{n,*})$, and compare this value with the true optimum $f_i^*(x_i^*)$ to obtain a measure of performance called relative Opportunity Cost (OC):

$$OC_i(n) = \frac{f_i^*(x_i^*) - f^*(x_i^{n,*})}{f_i^*(x_i^*)}.$$

OC values are always positive, and smaller values imply better performance. An OC of 0 means the predicted optimum design is as good as the true optimum. We then calculate the median OC values over all 100 simulations:

$$OC(n) = Median\{OC_i(n)\}_{i=1}^{100}.$$

This OC-curve describes the median performance of a policy as a function of the number of experiments.

**ISOMAP visualizations of parameter exploration**

To further characterize the exploration of the space, we projected the four-dimensional vectors representing specific synthesis conditions into a two-dimensional latent space using the Isomap algorithm [1], trained on the full set of experimental data collected in this study. Using this trained projection operator, we can plot the set of experiments performed, in a projected two-dimensional latent space, which is shown in **Figure S8**. The Isomap algorithm projects the four-dimensional specification of an experiment to a two-dimensional space in a way that attempts to preserve distances, allowing us to gauge the "spread" of the parameter space explored. In the plots, the original 6 data points are shown as red squares. The subsequent experiments suggested by the UCB policy are shown as blue circles. In addition to visualizing the actual experiments run during the validation phase, which is labeled as "Data" in **Figure S8**, we also include similar plots of experiments selected in simulated experimental campaigns based on different decision-making policies. These are shown in **Figure S8**, in the plots labeled "EI", "UCB", and "MV", each showing a typical distribution of simulated experimental choices. From visual inspection, we see that the validation campaign data explore more effectively than simulated EI and UCB policies, but not as much as the MV policy. The MV policy is not one for optimization - rather it is one mean to reduce global uncertainty, and hence will choose to explore the space more than exploit.

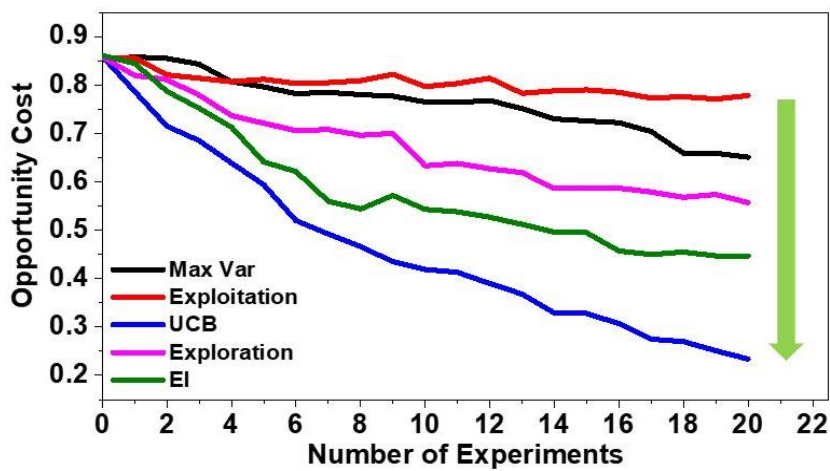

**Figure S1.** Simulation results based on the validation campaign. These plots show a measure of policy performance, OC, as a function of the number of simulated experiments. Of the five policies evaluated, the UCB policy had the best simulated performance.

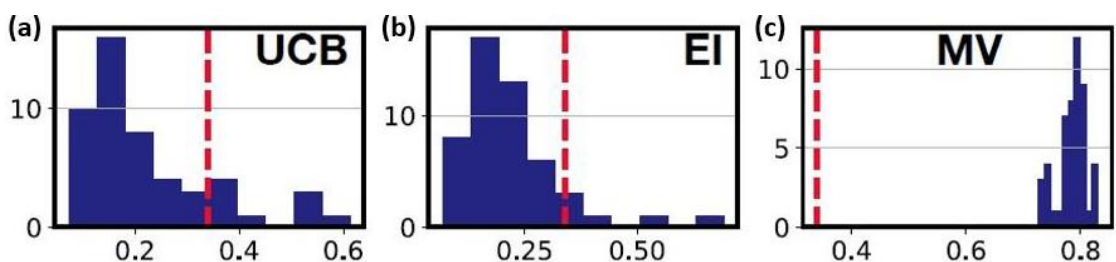

**Figure S2.** Mean distance from centroid distribution under simulated campaigns using the (a) UCB, (b) EI and (c) MV. The dashed red line indicates the actual distance (0.34) calculated from the physical experimental campaign. We observe that the distance from centroid is larger than a majority of those obtained under simulated UCB and EI policies, while lower than those obtained under the simulated MV policy, suggesting that the physical campaign robustly explored space to look for high performing synthesis conditions.

**Table S1.** The summarization of the synthesis conditions from prior knowledge and Bayesian Optimization predicted conditions, and their corresponding overpotential and Tafel slope.

| Sample | Mo Precursor (mol/L) | S Precursor (mol/L) | Mo / S Ratio | Temp (°C) | Time (h) | η10 (mV) | Tafel Slope (mV/dec) |
|---|---|---|---|---|---|---|---|
| $MoS_2$-1 | 0.03 | 0.9 | 0.033 | 205 | 26 | 288 | 110 |
| $MoS_2$-2 | 0.028 | 0.85 | 0.0329 | 200 | 24 | 243 | 84 |
| $MoS_2$-3 | 0.026 | 0.8 | 0.0325 | 195 | 22 | 240 | 70 |
| $MoS_2$-4 | 0.024 | 0.75 | 0.032 | 190 | 20 | 254 | 64 |
| $MoS_2$-5 | 0.022 | 0.7 | 0.0314 | 185 | 18 | 254 | 64 |
| $MoS_2$-6 | 0.02 | 0.65 | 0.0308 | 180 | 16 | 259 | 65 |
| $MoS_2$-P1 | 0.02 | 1.1 | 0.0182 | 180 | 36 | 249 | 82 |
| $MoS_2$-P2 | 0.038 | 0.65 | 0.0585 | 180 | 36 | 257 | 64 |
| $MoS_2$-P3 | 0.02 | 1.1 | 0.0182 | 189 | 33.3 | 219 | 88 |
| $MoS_2$-P4 | 0.02874 | 0.85729 | 0.0335 | 193.964 | 24.8 | 257 | 69 |
| $MoS_2$-P5 | 0.028529 | 0.7581 | 0.03763 | 190.017 | 25.09 | 259 | 66 |
| $MoS_2$-P6 | 0.02377 | 0.8619 | 0.02758 | 191.35 | 25.16 | 255 | 60 |
| $MoS_2$-P7 | 0.0285369 | 0.75067 | 0.03801 | 190.77 | 25.08 | 259 | 68 |
| $MoS_2$-P8 | 0.02435 | 0.87569 | 0.02780 | 198.68 | 19.08 | 254 | 79 |
| $MoS_2$-P9 | 0.02921 | 0.78856 | 0.03704 | 188 | 23.76 | 253 | 72 |
| $MoS_2$-Opt | 0.026072 | 0.801818 | 0.032516 | 195 | 22.07 | 243 | 64 |

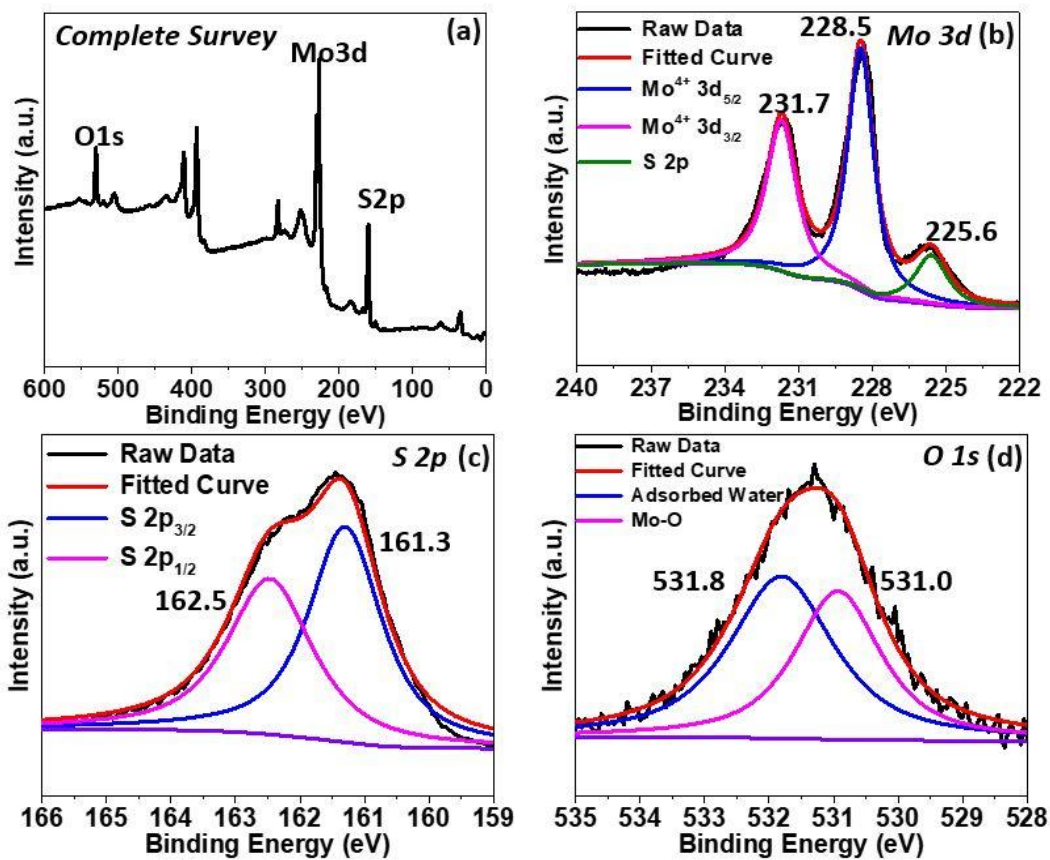

**Figure S3.** The deconvoluted XPS spectrums for MoS$_2$-P3.

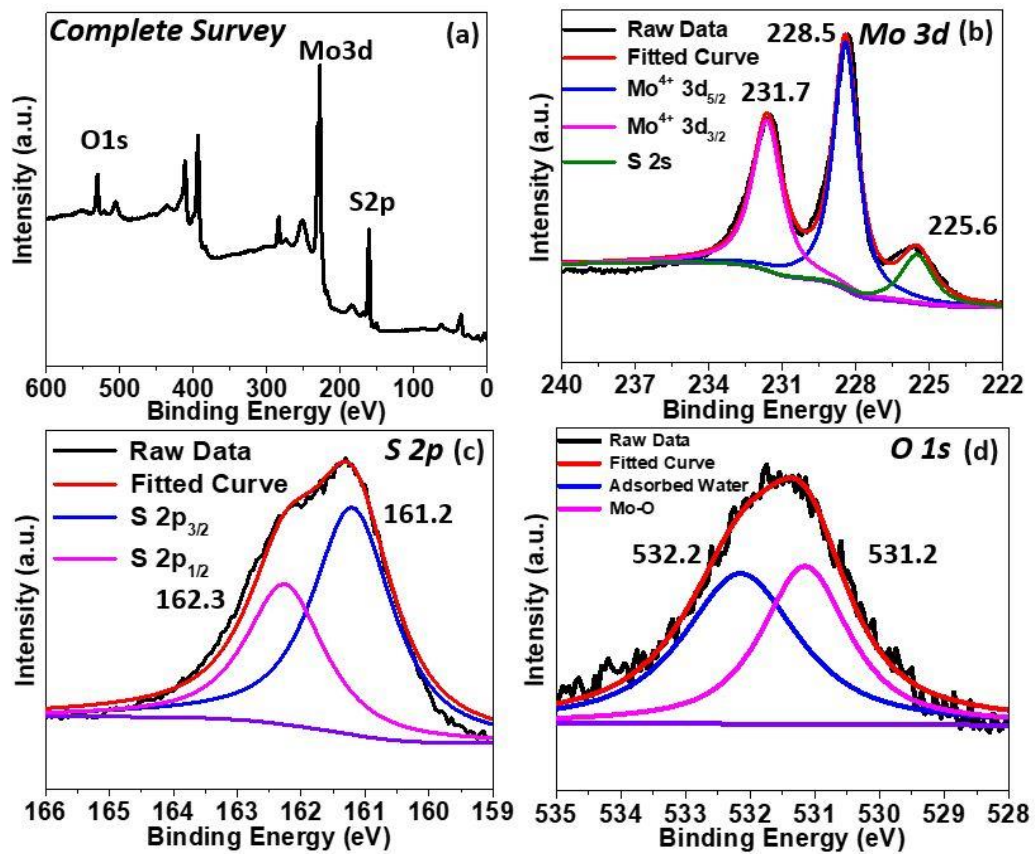

**Figure S4.** The deconvoluted XPS spectrums for MoS$_2$-P9.

**Table S2.** The summarization of the XRD extracted (002) peak position and interlayer spacing.

| Sample | (002) Peak Position (Degree) | (002) Direction Interlayer Spacing (nm) |
|---|---|---|
| $MoS_2$-P1 | 14 | 0.64 |
| $MoS_2$-P2 | 13.46 | 0.66 |
| $MoS_2$-P3 | 13.46 | 0.66 |
| $MoS_2$-P4 | 14 | 0.64 |
| $MoS_2$-P5 | 13.9 | 0.64 |
| $MoS_2$-P6 | 13.72 | 0.66 |
| $MoS_2$-P7 | 13.78 | 0.64 |
| $MoS_2$-P8 | 14.08 | 0.64 |
| $MoS_2$-P9 | 9.02 | 0.98 |
| $MoS_2$-Opt | 9 | 0.98 |

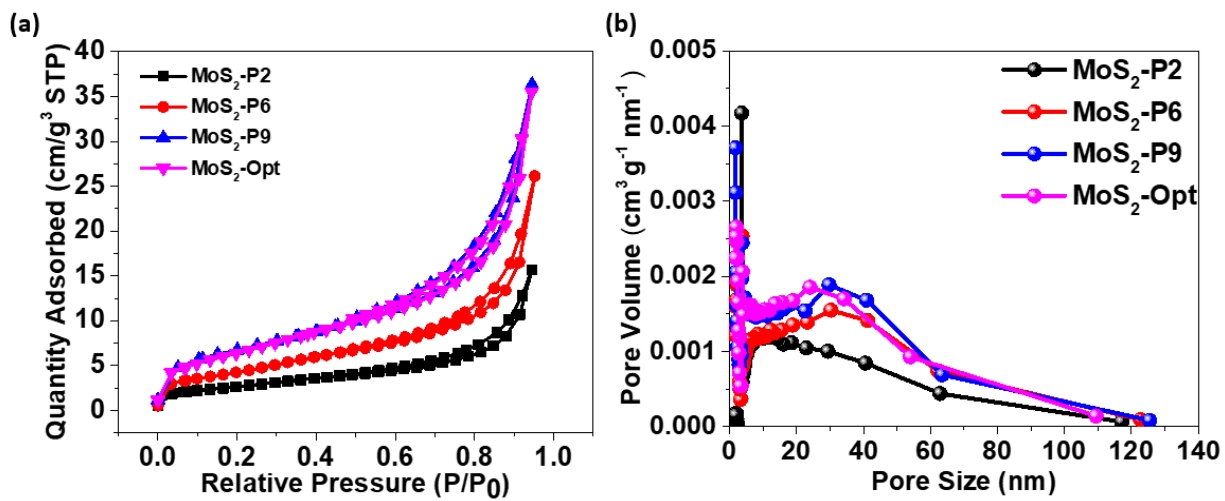

**Figure S5.** (a) The Nitrogen adsorption/desorption curve and (b) the pore size distribution of selected $MoS_2$ samples.

**Table S3.** The summarization of the BET surface area of selected samples.

| Sample | BET Surface Area (cm$^2$ g$^{-1}$) |
|---|---|
| MoS$_2$-P2 | 9.8 |
| MoS$_2$-P6 | 16.1 |
| MoS$_2$-P9 | 23.8 |
| MoS$_2$-Opt | 23.9 |

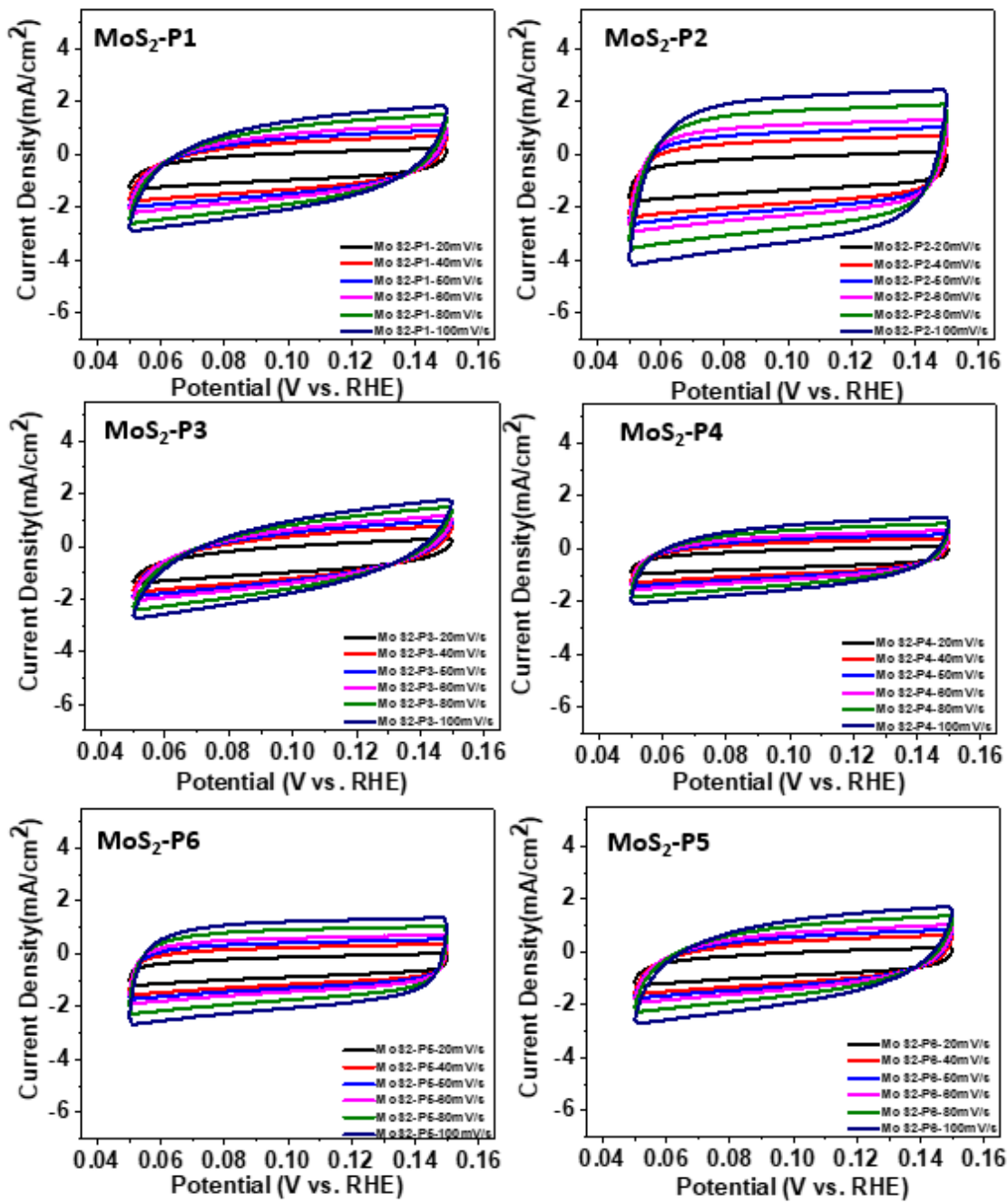

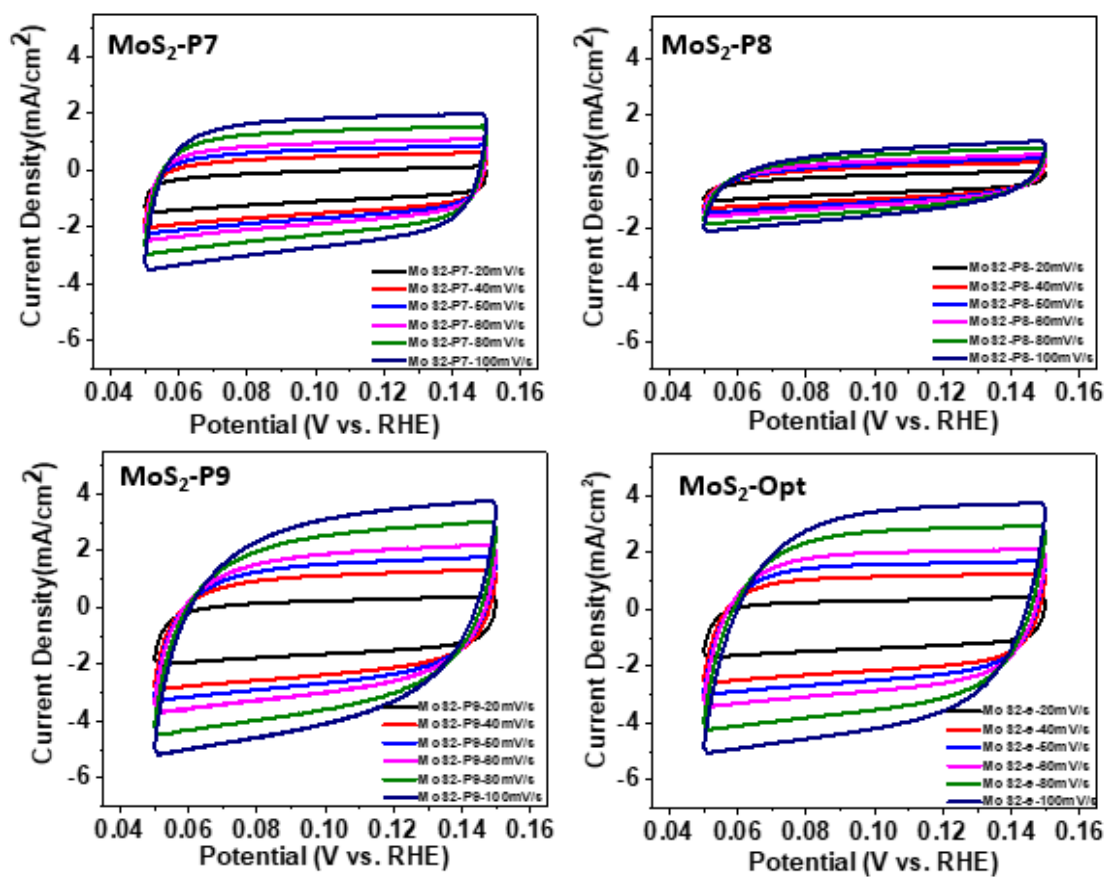

**Figure S6.** Cyclic voltammetry curves for all the sample which were used for extraction of double layer capacitance and ECSA.

**Table S4. The summarization of the $C_{dl}$ and ESCA for all samples.**

| Sample | $C_{dl}$ (mF cm$^{-2}$) | ECSA(cm$^2$) |
|---|---|---|
| MoS$_2$-P1 | 14.22 | 355.5 |
| MoS$_2$-P2 | 26.36 | 659 |
| MoS$_2$-P3 | 10.74 | 268.4 |
| MoS$_2$-P4 | 11.03 | 275.8 |
| MoS$_2$-P5 | 15.79 | 394.6 |
| MoS$_2$-P6 | 13.71 | 342.6 |
| MoS$_2$-P7 | 21.39 | 534.8 |
| MoS$_2$-P8 | 10.48 | 262 |
| MoS$_2$-P9 | 33.03 | 825.6 |
| MoS$_2$-Opt | 37.17 | 929.1 |

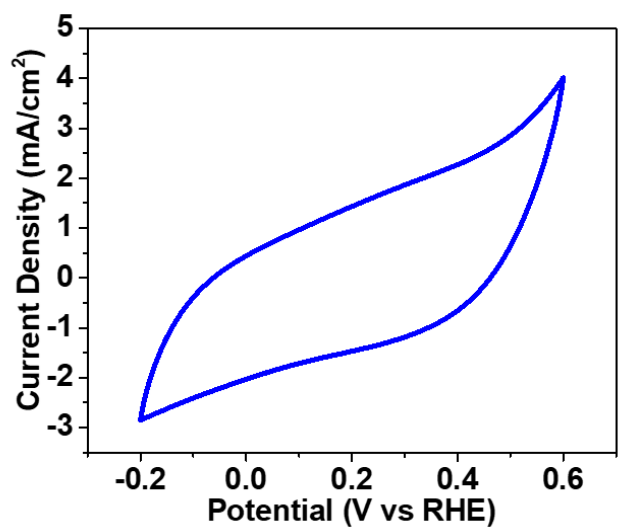

**Figure S7.** Cyclic voltammetry curves for MoS$_2$-Opt in Phosphate buffer at a scan rate of 50mV/s for TOF calculation.

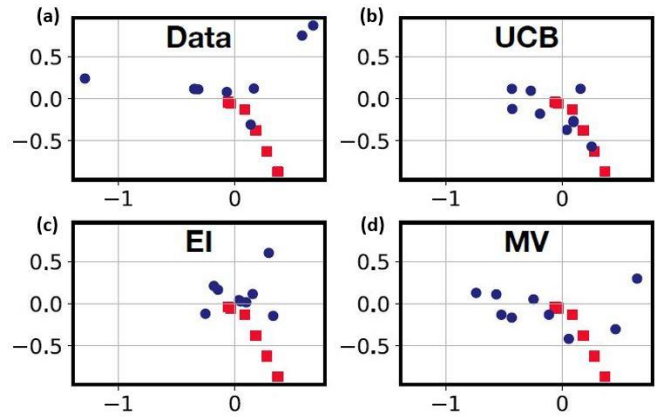

**Figure S8.** The exploration visualization for (a) real data, (b) UCB simulated data, (c) EI simulated data and (d) MV simulated data. Synthesis conditions are projected into a two-dimensional latent space using the Isomap method. Red points are the projection of the original data set, while blue points are additional experiments considered in the real campaign (a) or simulated campaigns (b-c) under the indicated decision-making policy.